\documentclass[journal]{IEEEtran}
\usepackage{epstopdf}
\usepackage{subfig}
\usepackage[utf8]{inputenc}
\usepackage{graphicx}
\usepackage{epstopdf}
\usepackage{amsmath}
\usepackage{amsfonts}
\usepackage{amssymb, color, bm}

\usepackage{mathtools}

\usepackage{tikz}
\usetikzlibrary{quantikz}

\usepackage{physics}

\usepackage{algorithmicx}
\usepackage[noend]{algpseudocode}
\usepackage{algorithm,setspace}
\usepackage[T1]{fontenc}

\usepackage{multicol}
\usepackage{lipsum}
\usepackage{mwe}

\DeclareMathOperator{\E}{\mathbb{E}}

\usepackage[utf8]{inputenc}
\allowdisplaybreaks
\usepackage{cite}

\usepackage{algorithm}
\usepackage[noend]{algpseudocode}

\usepackage{lineno}
\begin{document}

\title{Training Hybrid Classical-Quantum Classifiers via Stochastic Variational Optimization }
\author{Ivana Nikoloska,~\IEEEmembership{Graduate Student Member,~IEEE,}
        and~Osvaldo Simeone,~\IEEEmembership{Fellow,~IEEE.}
\thanks{I. Nikoloska and O. Simeone are with KCLIP, CTR, Department
of King's College London. e-mails: \{ivana.nikoloska, osvaldo.simeone\}@kcl.ac.uk. This work was supported by the European Research Council (ERC) under the European Union’s Horizon 2020 Research and Innovation Program (Grant Agreement No. 725731).}
}
\maketitle

\begin{abstract}

Quantum machine learning has emerged as a potential practical application of near-term quantum devices. In this work, we study a two-layer hybrid classical-quantum classifier in which a first layer of quantum stochastic neurons implementing generalized linear models (QGLMs) is followed by a second classical combining layer. The input to the first, hidden, layer is obtained via amplitude encoding in order to leverage the exponential size of the fan-in of the quantum neurons in the number of qubits per neuron. To facilitate implementation of the QGLMs, all weights and activations are binary. 
While the state of the art on training strategies for this class of models is limited to exhaustive search and single-neuron perceptron-like bit-flip strategies, this  letter introduces a stochastic variational optimization approach that enables the joint training of quantum and classical layers  via stochastic gradient descent. Experiments show the advantages of the approach for a variety of activation functions implemented by QGLM neurons.



\end{abstract}

\section{Introduction}
 Ever since Richard Feynman proposed the concept of quantum computers almost half a century ago, technology giants, startups and academic labs alike have competed against, and collaborated with each other, eager to make it a reality. Progress has had to contend with the stark limitations of current noisy-intermediate scale quantum (NISQ) systems providing 50-100 non-fault-tolerant qubits \cite{preskill2018quantum}. An emerging  potential practical use of NISQ hardware is quantum machine learning, a hybrid research discipline that combines machine learning and quantum computing \cite{biamonte2017quantum,havlivcek2019supervised,schuld2021machine}. 
 
 Many classical techniques, ranging from kernel methods \cite{hofmann2008kernel} and Boltzmann machines \cite{hinton2007boltzmann} to deep learning models like convolutional \cite{krizhevsky2012imagenet} and graph neural networks \cite{kipf2016semi}, now have quantum counterparts \cite{mengoni2019kernel,amin2018quantum,cong2019quantum,verdon2019quantum}, which can operate, at least on a small scale, on NISQ hardware. These methods apply classical optimization routines to select parameters that define the operation of a quantum circuit. Such parameters are most often continuous and optimized via gradient descent techniques that apply the so-called parameter-shift rule, an exact form of finite-difference differentiation \cite{schuld2021machine}. Alternative approaches, which may be more promising in the short term, involve hybrid quantum-classical models, where classical computation, e.g., for feature extraction, is combined with quantum parametric circuits  \cite{mari2020transfer}. 
 
 \begin{figure}[tbp]
\centering
\includegraphics[width=1\linewidth]{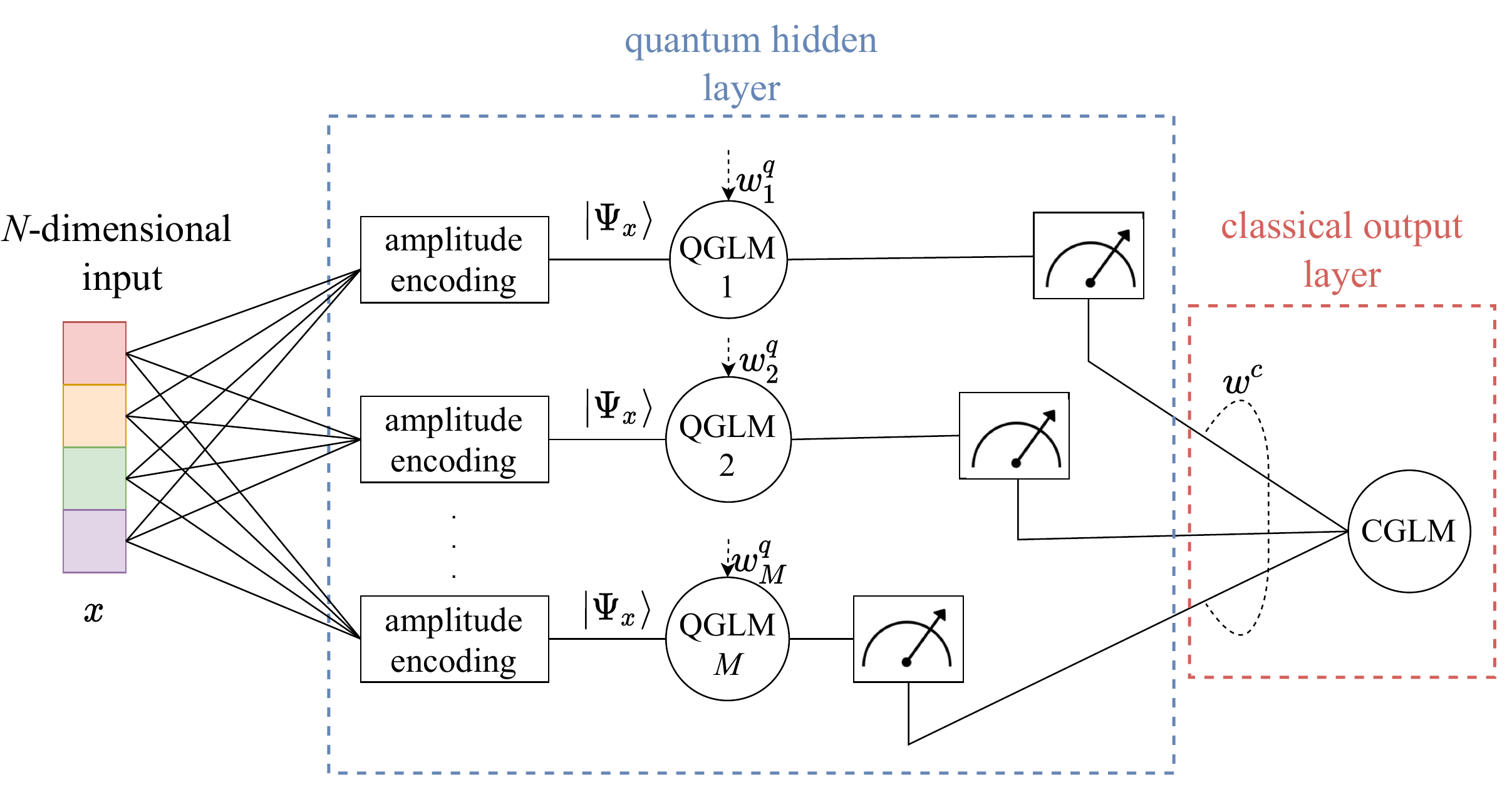}
\caption{In the studied hybrid classical-quantum classifier, a quantum hidden layer, fed via amplitude encoding and consisting of quantum generalized linear models (QGLMs), is followed by a classical combining output layer with a single classical GLM (CGLM) neuron. All weights and activations are binary.  
}
\vspace*{-3.7mm}
\label{model}
\end{figure}
 
 A notable example of hybrid quantum-classical models is the multi-layer artificial neural network introduced in \cite{tacchino2019artificial,tacchino2020quantum}, which includes stochastic binary neurons implementing generalized linear models (GLMs). The key merits of this architecture are that: (\emph{i}) quantum implementations of GLMs, referred to as QGLM, have an exponentially large number of inputs and number of synaptic weights in the number of physical resources -- qubits -- within the neuron; and (\emph{ii}) unlike more conventional deterministic models, such as \cite{arthur2022hybrid}, the outcomes of the quantum neurons need not be averaged by performing multiple measurements to obtain the expectations of given observables. 
 
 The  QGLM-based neural network considered in \cite{tacchino2019artificial,tacchino2020quantum} has binary synaptic weights, as well as binary, classical, activations, rendering standard optimization methods infeasible. For models comprised of one or two QGLM neurons, the weight vectors were optimized in \cite{tacchino2019artificial,tacchino2020quantum} via a single-neuron perceptron-like sign-flips strategy or via exhaustive search. Both approaches are inapplicable or infeasible for models comprised of an arbitrary number of neurons. 
 
 In this paper, we address this issue by focusing -- as in the experiments presented in \cite{tacchino2020quantum}  --  on the hybrid classical-quantum two-layer architecture illustrated in Fig. 1. In it,  a first layer of QGLMs is followed by a second classical combining layer. The input to the first, hidden, layer is obtained via amplitude encoding (see, e.g., \cite{schuld2021machine}). This letter introduces a stochastic variational optimization (SVO) approach \cite{bird2018stochastic} that enables the joint training of quantum and classical layers  via stochastic gradient descent. The proposed SVO-based training strategy operates in a relaxed continuous space of variational classical parameters. 
 Experiments via computer simulations show the advantages of the approach for a variety of activation functions implemented by the QGLM neurons.

\vspace*{-3.7mm}

\section{Model}
In this section, we describe the hybrid classical-quantum classifier depicted in Fig. 1.

\vspace*{-3.7mm}
\subsection{Hybrid Classical-Quantum Binary Classifier}

As illustrated in Fig. 1, we consider a two-layer architecture implementing a binary classifier on an input, classical, vector $x$. The first, hidden, layer consists of $M$ QGLM-based neurons, and the last layer of a classical GLM (CGLM) neuron producing the final classification decision. 

The input vector $x=[x_0,...,x_{N-1}]^{T}$, which is assumed to be binary with $x_n\in\{-1,+1\}$, is mapped to the amplitude vector of a pure quantum state consisting of $\log_2(N)$ qubits (assumed to be an integer) as \begin{align}\label{psi_x}
    \ket{\Psi_x} = \frac{1}{\sqrt{N}} \sum_{n=0}^{N-1} x_n \ket{n}, 
\end{align}where $\ket{n}$, with  $n\in \{0,...,N-1\}$, denotes the $n$th vector of the computational basis (see, e.g., \cite{mermin2007quantum}). We note that preparing the quantum state \eqref{psi_x} from a general binary input vector $x$ entails a minimal complexity of $N/\log_2(N)$ computational steps, most of which are two-qubit gates, unless vector $x$ has a specific structure (e.g., it is a sample from a smooth probability distribution) \cite{schuld2021machine}.

As seen in Fig. 1, the quantum state $\ket{\Psi_x}$ is prepared $M$ times and input to each of the QGLM neurons in the first layer. Each $m$th QGLM neuron produces a stochastic binary output $y_m \in \{-1,+1\}$ as a function of the input state $\ket{\Psi_x}$ and of an $N$-dimensional vector $w^q_m=[w^q_{m,0},...,w^q_{m,N-1}]^{T}$ of binary weights, with $w^q_{m,n} \in \{-1,+1\}$. 

The $M$ binary digits $y=[y_1,...,y_M]^{T}$ produced by the first layer are then fed to a CGLM neuron, which outputs a final, stochastic, decision $z\in\{-1,+1\}$ for binary classification as a function of its own vector of binary weights  $w^c=[w^c_1,...,w^c_M]^{T}$.

We emphasize that, unlike deterministic models such as \cite{arthur2022hybrid}, the outputs of the quantum neurons need not be averaged to obtain the expectation of an observable over multiple runs. Rather, the outcomes of the quantum neurons are obtained via a single measurement, producing a stochastic binary output.


\vspace*{-3.7mm}
\subsection{Classical GLM (CGLM) Neuron}
In order to describe the operation of the neurons, let us start with the classifying CGLM neuron in the last layer. As seen, the CGLM neuron receives an $M$-dimensional binary input vector $y$ and produces a final binary decision $z\in\{-1,+1\}$. The probability of the binary output given the input vector $y$ is given by 
\begin{align}\label{MPneuron}
p_{w^c}(z = 1 \, | y) = \textrm{g}^{c} \left(\sum_{m=1}^{M} w^c_m y_m\right) = \textrm{g}^{c} \left(y^{T}w^{c}\right),
\end{align}
where $\textrm{g}^{c}(\cdot)$ denotes the  response function of the CGLM and $(\cdot)^{T}$ denotes transposition. Typical examples of the response function $\textrm{g}^{c}(\cdot)$, which must be invertible, of GLMs include the sigmoid function and the Gaussian cumulative distribution function (see, e.g., \cite{simeone2018brief}).

\vspace*{-3.7mm}
\subsection{Quantum GLM (QGLM) Neuron}
Quantum procedures closely mimicking the operation  \eqref{MPneuron}
of the CGLM neuron can be designed using (approximately) $\log_2(N)$ qubits (see \cite{schuld2021machine} for an overview). The fact that $\log_2(N)$ qubits are sufficient to process an exponentially larger input $x$, of size $N$, illustrates the potential computational advantage of quantum information processing in this context. We refer to quantum circuits implementing the stochastic mapping defined by the conditional distribution
\begin{align}\label{QGLMneuron}
p_{w^q_m}(y_m = 1 \, | x) = \textrm{g}^{q} \left(\sum_{n=0}^{N-1} w^q_{m,n} x_n\right) = \textrm{g}^{q} \left(x^{T} w_{m}^{q}\right),
\end{align} for some (invertible) response function $\textrm{g}^{q}(\cdot)$, given binary weight vector $w^q_m$, as QGLM neurons. The outputs of the $M$ QGLMs are independent given the input $x$, and hence we have $p_{w^q} ( y \, | \, x ) = \prod_{m=1}^{M} p_{w^q_m}(y_m  \, | x)$, where $w^q=\{w^q_m\}_{m=1}^{M}$.

Several implementations of QGLM neurons have been proposed in the literature using different quantum circuits. The main goal of these circuits is to produce a stochastic binary output with probabilities which are a function of the inner product $\bra{\Psi_{w^q}}\ket{\Psi_x} = x^{T} w_{m}^{q} $ 
between the input state $\ket{\Psi_x}$ and the amplitude-encoded binary weight vector $\ket{\Psi_{w^q}} = 1/\sqrt{N}\sum_{n=0}^{N-1} w^q_n \ket{n}$. Different solutions, along with the resulting response functions are listed in Table I, accompanied by relevant references. In the experiments provided in Sec. IV we consider all these options.





\begin{table*}[!h]
\centering
\caption{List of response functions for QGLM neurons}
\label{table3}
\begin{tabular}{|l|l|l|}
\hline
Name & $\textrm{g} \left(\cdot \right)$  & Routine        \\ \hline
Quadratic (Q)      & $\textrm{g}(x^{T} w_{m}^{q}) = |x^{T} w_{m}^{q}|^2$ & Sign-flip blocks \cite{tacchino2019artificial}, \cite{tacchino2020quantum}    \\ \hline
Biased quadratic (BQ)      & $\textrm{g}(x^{T} w_{m}^{q}) = 1/2 + 1/2|x^{T} w_{m}^{q}|^2$   & Swap test \cite{kobayashi2003quantum}  \\ \hline
Biased centered quadratic (BCQ)	       & $\textrm{g}(x^{T} w_{m}^{q}) = 1/2 + 1/2|x^{T} w_{m}^{q} - 1/2|^2$ &Swap test on extra dimensions \cite{zhao2019quantum}  \\ \hline
Linear (L)  & $\textrm{g}(x^{T} w_{m}^{q}) = 1/2 + 1/2 x^{T} w_{m}^{q}$ & Interference circuit \cite{schuld2017implementing}  \\ \hline

\end{tabular}
\end{table*}

\vspace*{-3.7mm}

\section{Stochastic Variational Optimization-Based Training}

In this section, we first define the problem of training the hybrid classical-quantum classifier defined in the previous section, and then introduce an SVO-based training procedure. 

%


\vspace*{-3.7mm}
\subsection{Problem Definition}

We assume to have access to a training set $\mathcal{D}$ of input-output samples $(x,z)$, with $x \in \{-1,1\}^N$ and $z \in \{-1,1\}$. Given the model described in the previous section, we define the log-loss on an example $(x,z)$ as
\begin{align}\label{loss_overall}
    \mathcal{L}_{(x,z)} (w)  &= - \ln \left( \E_{p_{w^q} \left( y \, | \, x \right)} \left[p_{w^c} \left( z \, | \, y \right)\right]\right),
\end{align}
which is a function of the weights $w=\{\{w^q_m\}_{m=1}^{M},w^c\}$ of the $M$ QGLMs and of the CGLM. Note that computing the log-loss requires averaging over the outputs $y$ of the hidden QGLMs. Using Jensen's inequality, the loss in \eqref{loss_overall} can be bounded as
\begin{align}\label{loss_overall_bound}
    \mathcal{L}_{(x,z)} (w) &\leq  \E_{p_{w^q} \left( y \, | \, x  \right)} \left[ \ell \left(z, \textrm{g} (y^{T}w^{c})\right) \right] \coloneqq L_{(x,z)} (w),
\end{align} where we have defined the negative log-probability of the output of the CGLM as
\begin{align}\label{chain}
    -\ln p_{w^c} ( z \, | \, y) &= \ell \left(z, \textrm{g} (y^{T}w^c)\right),
\end{align}
with $\ell (a,b) = - \ln (b^{\frac{1+a}{2}} (1-b)^{\frac{1-a}{2}} )$.

The training objective is to minimize the loss bound in \eqref{loss_overall_bound} over the $NM+M$ binary weights $w$ by addressing the problem\begin{align}\label{loss_prob}
    \underset{w\in\{0,1\}^{NM+M}}{\text{min}} \,\,\,  \sum_{(x,z) \in \mathcal{D}} L_{(x,z)} (w).
\end{align}

\vspace*{-3.7mm}
\subsection{SVO-Based Training Algorithm}

The direct optimization of problem (\ref{loss_prob}) is problematic due to the discrete nature of its domain.  In this paper, we propose to address this problem via SVO. SVO 
is a generalization of stochastic optimization via simultaneous perturbation \cite{spall1997one} that has the key merit of being  applicable also to discrete models. It is noted that stochastic optimization via simultaneous perturbation  was found to be effective in \cite{havlivcek2019supervised} to optimize a parameterized quantum circuit with continuous model parameters. 

SVO makes use of the observation that the minimum of a collection of values cannot be larger than an arbitrary average of such values, i.e., \cite{bird2018stochastic}
\begin{align}\label{loss_prob_bound}
    &\underset{w}{\text{min}} \,\,\,  \sum_{(x,z) \in \mathcal{D}} L_{(x,z)} (w) \nonumber\\
    \leq &\underset{q(w \, | \, \phi)}{\text{min}} \,\,\, \E_{q(w \, | \, \phi)} \left[ \sum_{(x,z) \in \mathcal{D}} L_{(x,z)} (w) \right],
\end{align}
where $q(w \, | \, \phi)$ represents a parametric distribution on the space of the model parameters $w$ that depends on a continuous-valued vector of parameters $\phi$. Furthermore, if the family of distributions is sufficiently large to include all  distributions concentrated at the possible values of $w$, the relation in (\ref{loss_prob_bound}) holds with equality.

Given that the model parameters are discrete, we define the variational distribution $q(w \, | \, \phi)$ as the product-Bernoulli probability mass function
\begin{align}\label{search_distr}
    q(w \, | \, \phi) = & \prod_{m=1}^{M} \prod_{n=0}^{N-1}  \sigma(\phi_{m,n}^q)^{\frac{1+w_{m,n}^q}{2}} (1-\sigma(\phi_{m,n}^q))^{\frac{1-w_{m,n}^q}{2}} \nonumber \\ & \times \prod_{m=1}^{M} \sigma(\phi_{m}^c)^{\frac{1+w_{m}^c}{2}} (1-\sigma(\phi_{m}^c))^{\frac{1-w_{m}^c}{2}}
    ,
\end{align}
where $\sigma(a)=(1+\exp(-a))$ denotes the sigmoid function and we have defined the $NM+M$ variational parameters $\phi=\{\{\{\phi^q_{m,n}\}_{m=1}^{M}\}_{n=0}^{N-1},\{\phi^c_{m}\}_{m=1}^{M}\}$. These are the natural parameters defining the probabilities that the corresponding weights equal $1$. Specifically, the real parameter $\phi^q_{m,n}$ yields the probability $\sigma(\phi^q_{m,n})$ that we have  $w^q_{m,n}=1$, and $\phi^c_{m}$ gives the probability $\sigma(\phi^c_{m})$ of event $w^c_{m}=1$ under the variational distribution $q(w \, | \, \phi)$. With this choice, we address problem (\ref{loss_prob}) by minimizing the upper bound in \eqref{loss_prob_bound} over the $NM+M$-dimensional real vector $\phi$. 


To this end, we aim at approximating the gradient-based update\begin{align}\label{phi_exp}
    \phi &\leftarrow \phi+ \eta \nabla_{\phi} \E_{q(w \, | \, \phi)} \left[ \sum_{(x,z) \in \mathcal{D}} L_{(x,z)} (w) \right] \nonumber \\ & = \phi + \eta \E_{q(w \, | \, \phi)} \left[ \left(\sum_{(x,z) \in \mathcal{D}} L_{(x,z)} (w) - b \right) \nabla_\phi \ln q(w \, | \, \phi) \right],
\end{align}
where $\eta>0$ is the learning rate and $b$ is an arbitrary vector of the same dimensions as $\phi$. The equality in \eqref{phi_exp} follows from the standard REINFORCE expression \cite{mohamed2020monte}. The gradient of the log-variational distribution in (\ref{phi_exp}) has the simple form
\begin{align}\label{grad}
    \nabla_\phi \ln q(w \, | \, \phi) = \frac{1+w}{2} - \sigma(\phi),
\end{align}where the functions are evaluated entry-wise. 

In order to obtain a practical update rule, the expectation over the variational distribution $q(w|\phi)$ and the sum over the data set in \eqref{phi_exp}  are estimated using samples from both the variational distribution $q(w|\phi)$ and from the data set $\mathcal{D}$. With a single sample $w\sim q(w|\phi)$, we obtain the doubly stochastic estimate \cite{mohamed2020monte}\begin{align}\label{phi_MC}
    \phi \leftarrow \phi + \eta  \left[ \left(\frac{|\mathcal{D}|}{|\mathcal{D}_b|}\sum_{(x,z) \in \mathcal{D}_b} L_{(x,z)} (w) - b \right) \nabla_\phi \ln q(w \, | \, \phi) \right],
\end{align}where $|\cdot|$ is the cardinality of the argument set and $\mathcal{D}_b$ represents a mini-batch of the data set $\mathcal{D}$. The estimate \eqref{phi_MC} can be readily extended to any number of samples from $q(w|\phi)$.
Importantly, the update \eqref{phi_MC} is local: Each entry of vector $\phi$ can be updated separately based on knowledge of the global loss $L_{(x,z)}(w)$ for the batch of samples $\mathcal{D}_b$. In fact, the gradient \eqref{grad} can be computed entry-wise and hence separately for each parameter in vector $\phi$. 

The update \eqref{phi_MC} can be interpreted in a manner similar to stochastic optimization via simultaneous perturbations \cite{spall1997one} (which is also related to evolution-based strategies \cite{JMLR:v15:wierstra14a}). At each iteration, the random sampling of the weights from the variational distribution $q(w|\phi)$ explores the space of the binary model parameters ``around'' the current variational probability vector $\phi$.


It is important to choose a baseline vector $b$ in an adaptive manner so as to minimize the variance of the gradient estimate \cite{mohamed2020monte}. This can be done as derived in (\ref{phi_exp}) as \cite{peters2008reinforcement}
\begin{align}\label{baseline}
    b = \frac{\E \left[ \left( \sum_{(x,z) \in \mathcal{D}} L_{(x,z)} (w) \nabla_\phi \ln q(w \, | \, \phi) \right)^2 \right]}{\E \left[ \left( \nabla_\phi \ln q(w \, | \, \phi) \right)^2 \right]},
\end{align}where again operations are carried out entry-wise. In order to estimate the expectations in \eqref{baseline}, we use two moving averages, estimated using an exponentially decaying factor $\gamma$, one for the numerator and one for the denominator in \eqref{baseline} across the iterations \cite{peters2008reinforcement}. Note that the baseline can also be updated locally and separately for each local parameter.



Moreover, training the model using the local update rule does not require estimating the response functions, i.e., the expectation of the outcomes of the quantum neurons. For quantum hardware, one can only take a finite number of measurements, so one can never determine a circuit’s expectation values exactly. This makes training the proposed architecture less computationally intensive than variational quantum circuits relying on parameters shift rules that require taking the difference of two circuit expectation values, with forward and backward shifts in angles \cite{schuld2017implementing}, \cite{rivera2021avoiding},\cite{arthur2022hybrid}.

\vspace*{-3.7mm}

\section{Experiments}
In this section, we provide experimental results to elaborate on the performance of the proposed training scheme.
\vspace*{-3.7mm}
\subsection{Data Set}
We consider a prototypical image data set, namely Bars-and-Stripes (BAS) which is widely used in related studies \cite{tacchino2019artificial, tacchino2020quantum, arthur2022hybrid, benedetti2019generative, li2020limitations}. Each sample of size $d \times d$ illustrates bars, stripes, or random patterns. The samples are transformed from $d \times d$ pixel images to binary strings $x$ of length $d^2$. As in \cite{tacchino2019artificial, tacchino2020quantum, benedetti2019generative, li2020limitations}, we consider $d = 4$. The samples containing bars or stripes are labeled as $z = 1$, whilst the random patterns are labeled as $z = -1$. 


\vspace*{-3.7mm}
\subsection{Schemes and Benchmarks}
We compare the performance of the SVO-based strategy with the perception-like scheme proposed in \cite{tacchino2019artificial} to train single quantum neurons. To adapt the scheme in \cite{tacchino2019artificial} to the two-layer architecture studied in this paper, we train each neuron individually as in \cite{tacchino2019artificial}. A final classification decision is then made by the classical neuron in the second layer via a majority rule, i.e., a label is selected if it is chosen by at least half of the neurons. Specifically, following \cite{tacchino2019artificial}, the weight vector $w_m^q$ for QGLM $m$ is updated as follows. If the respective neuron classifies a sample with a negative label as positive, one flips a fixed portion $\eta\in[0,1]$ of the binary weights, which are selected at random among the positions at which the signs of the weights and the sample coincide. Conversely, if the respective neuron classifies a positive sample as negative, one flips the same portion of the weights, which are selected at random among the positions at which the signs of the weights and the sample differ.

\vspace*{-3.7mm}
\subsection{Model Architecture and Hyperparameters}
We consider the hybrid classical-quantum model in Fig. 1 with $M = 32$ hidden quantum neurons, and a single classical output neuron. We use $|\mathcal{D}| = 30$ randomly selected samples for training, and testing is carried out with the same number of independently generated samples. We use SGD with $4000$ training iterations. For the sign-flips scheme, the fraction of flipped signs is set to $\eta = 0.625$ (which was optimized numerically through exhaustive search). For SVO, the learning rate $\eta$ is altered based on a cyclical schedule as in  \cite{smith2017cyclical} as
\begin{align}
    &\eta= \eta_{\text{base}} + (\eta_{\text{max}}-\eta_{\text{base}}) \nonumber \\
    &\times \text{max}(0, (1-t)) \times \left(1+\sin(\frac{c \pi}{2})\right),
\end{align}
where $i$ denotes the current training iteration; $s=1000$ denotes the step size; the base and maximum values for the learning rate are set to $\eta_{\text{base}} = 0.1$, and $\eta_{\text{max}} = 0.9$, respectively; and we have defined $c = \left\lceil \left(1+i/(2s)\right) \right\rceil$ and $t = \left|i/s - 2 c + 1\right|$. In addition, for SVO, the batch size is set to $|\mathcal{D}^b| =16$, and the number of samples of the model parameters used to evaluate the expectations in \eqref{phi_MC} is set to $10$. We adopt a standard sigmoid for the response function $g^c(\cdot)$ of the CGLM.



\vspace*{-3.7mm}
\subsection{Results}
\begin{figure}[tbp]
\centering
\includegraphics[width=1\linewidth]{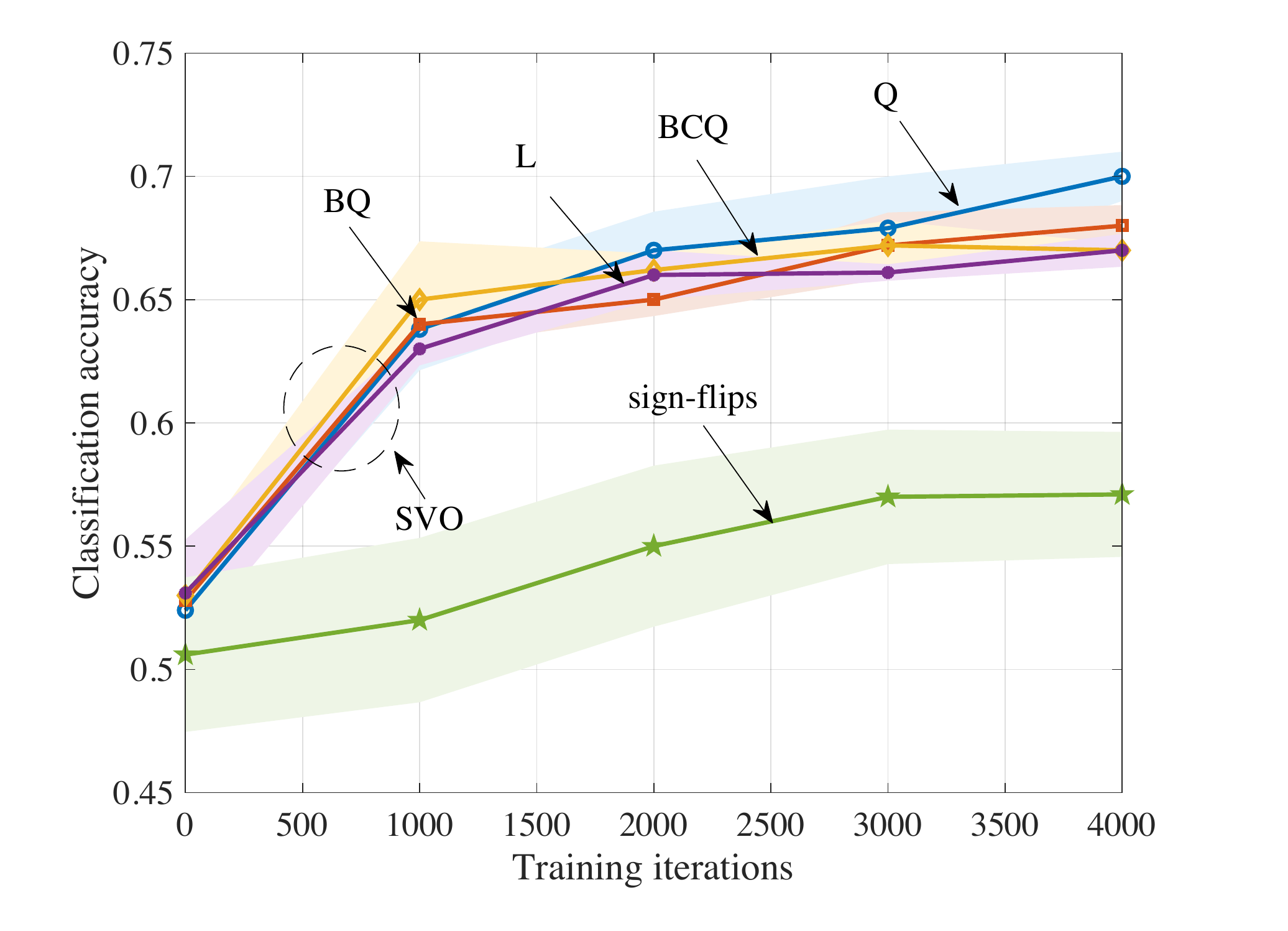}
\caption{Classification accuracy as a function of the training iteration for the benchmark sign-flips scheme \cite{tacchino2019artificial} and the proposed SVO-based procedure for the BAS data set. 
The results are averaged over $5$ independent trials.}
\vspace*{-3.7mm}
\label{LLR_horizontal}
\end{figure}

We plot the classification accuracy for all schemes and benchmarks in Fig.~\ref{LLR_horizontal} as a function of the training iterations. For SVO, we consider all response functions, while we illustrate only the quadratic (Q) response function for the sign-flips scheme, since it behaves similarly for all response functions. The sign-flips scheme \cite{tacchino2019artificial} is seen to improve as training proceeds, however, the accuracy saturates after around $3,000$ iterations. The reason is that there is no single hyperplane separating the bars and the stripes from the random patterns, making the task difficult to solve for individual neurons in the ensemble. Conversely, the proposed SVO scheme is seen to achieve higher classification accuracy for all response functions. In particular, the QGLM using the Quadratic (Q) response function yields fastest convergence and achieves the best performance. Due to the additional bias terms resulting from the swap test routine, the QGLMs relying on the Biased quadratic (BQ) and Biased centered quadratic (BCQ) response functions are slower to learn, but ultimately converge after around $3,000$ training iterations.

\vspace*{-3.7mm}
\section{Conclusion}
In this paper, we studied a two-layer hybrid classical-quantum classifier with binary weights and activations in which a first layer of QGLMs is followed by a second classical combining layer. For the considered class of models, we proposed a SVO-based training scheme that results in local learning rules for joint weight optimization of the quantum and classical layers via stochastic gradient descent. The proposed method can be naturally extended to architectures with multiple layers of QGLMs (see \cite{tacchino2020quantum}), although this is not elaborated on here due to the practical challenges associated with repeating amplitude encoding steps in between quantum layers.





\bibliographystyle{IEEEtran}
\bibliography{litdab.bib}

\end{document}